\documentclass[prl,twocolumn,showpacs,floatfix]{revtex4}

\usepackage{epsfig}
\usepackage{amssymb}

\newcommand\fig[1]{Fig.~\ref{#1}}
\newcommand\figN[1]{Fig.~#1}

\newcommand\eqn[1]{(\ref{#1})}
\newcommand\eqns[2]{(\ref{#1})-(\ref{#2})}

\newcommand{\dslash}{{\partial\!\!\!/}}
\newcommand{\ie}{, {\it i.e.}, }
\newcommand{\eg}{, {\it e.g.}, }

\begin{document}

\title{Condition for gapless color-$\bar{3}$ excitations in NJL models}

\author{F. Sandin}
\affiliation{Department of Physics, Lule{\aa} University of Technology,
SE-97187 Lule\aa , Sweden}

\author{A.M. \"Ozta\c{s}}
\affiliation{Department of Physics, Hacettepe University,
TR-06532 Ankara, Turkey}

\begin{abstract}
We present an exact condition for the existence of gapless quasiparticle excitations in NJL models of color superconducting quark matter with a quark-quark interaction in the scalar color-antitriplet channel. The condition can be represented by a rotated ellipse in the plane of mass and chemical potential differences for the paired quark fields.
\end{abstract}


% 12.38.Mh	Quark-gluon plasma
% 24.85.+p	Quarks, gluons, and QCD in nuclei and nuclear processes
% 25.75.Nq	Quark deconfinement, quark-gluon plasma production, and phase transitions

\maketitle

%%%%%%%%%%%%%%%%%%%%%%%%%%%%%%%%%%%%%%%%%%%%%%%%%%%%%%%%%%%%%%%%%%%

At high baryon density and low temperature matter is believed to be in a color superconducting state, which is characterized by condensates of quark Cooper pairs~\cite{Barrois:1977xd,Bailin:1983bm,Alford:1997zt,Rapp:1997zu}. A superconducting phase typically has an energy gap in the density of states, which corresponds to the lowest excitation energy of a quasiparticle pair. However, if the difference between the Fermi momenta of the paired quarks is sufficiently large, {\it gapless} quasiparticle excitations could exist~\cite{Shovkovy:2003uu,Huang:2003xd,Alford:2003fq,Alford:2004hz}. The presence of gapless phases could have observable consequences\eg the high specific heat and neutrino emissivity could affect the cooling behavior of compact stars~\cite{Alford:2004zr}. It has been found, however, that gapless phases might suffer from a chromomagnetic instability~\cite{Huang:2004am,Huang:2004bg,Casalbuoni:2004tb,Giannakis:2004pf,Fukushima:2005cm,Abuki:2005ms} and it is presently unclear whether gapless phases appear at temperatures relevant for compact star evolution~\cite{Alford:2004hz,Ruster:2005jc,Blaschke:2005uj,Abuki:2005ms}.
%Especially, the constituent strange quark mass and the strength of the quark-quark interactions have a great impact on the gapless phase structure.
It is therefore important to improve the understanding of gapless phases.
In this paper we derive an exact condition for the existence of gapless excitations in the frequently used Nambu--Jona-Lasinio (NJL) model of color superconducting quark matter. A qualitatively useful graphical representation of the condition, and some well-known approximations are also presented.

The most dense environment where quark matter is expected to exist is in the core of neutron stars, which are subject to a gravitational instability that limits the maximum density to $\sim 10^{15}$~g/cm$^3$~\cite{Kettner:1994zs}. This corresponds to a maximum quark number chemical potential of $\mu \sim 500$~MeV and a maximum baryon number density of $n_B\sim 10\,n_0$, where $n_0=0.17$~fm$^{-3}$ is the baryon number density in nuclear matter. Since the charm quark mass is higher than the maximum chemical potential, it is sufficient to consider up ($u$), down ($d$) and strange ($s$) quarks. The quark spinors are
\begin{equation}
q^T = \left(\psi_{ur}, \psi_{ug}, \psi_{ub}, \psi_{dr}, \psi_{dg}, \psi_{db}, \psi_{sr}, \psi_{sg}, \psi_{sb}\right),
\label{spinors}
\end{equation}
where $r$, $g$ and $b$ represent red, green and blue colors. The NJL model of superconducting quark matter is based on effective point-like four-fermion interactions and is described in\eg~\cite{Buballa:2003qv,Ruster:2005jc,Blaschke:2005uj,Abuki:2005ms}. Here we repeat some of the essential points. The Lagrangian density is
\begin{equation}
	{\cal L}_{\it eff} = {\bar q} (i \dslash - \hat{m} + \hat{\mu}\gamma^0) q
	+ {\cal L}_{\bar{q}q} + {\cal L}_{qq},
\label{lagrangian}
\end{equation}
where $\hat{m}=\text{diag}_f(m_u,\,m_d,\,m_s)$ is the current quark mass matrix in flavor space. ${\cal L}_{\bar{q}q}$ and ${\cal L}_{qq}$ are the effective interaction terms, which are used at mean-field level in Hartree approximation. Explicitely
\begin{eqnarray}
	{\cal L}_{\bar{q}q} &=& G_S \sum_{a=0}^8 \Big[({\bar q}\tau_a q)^2
	+ ({\bar q} i\gamma_5\tau_a q)^2\Big] \nonumber \\
	&-&K\left[\text{det}_f(\bar{q}(1+\gamma_5)q)+\text{det}_f(\bar{q}(1-\gamma_5)q)\right],
\label{qbarqlag} \\
	{\cal L}_{qq} &=& G_D\!\!\!\!\!\!\sum_{a,b = 2,5,7}\!\!\!\!\!\! %\sum_{b = 2,5,7}
	(\bar{q} i\gamma_5 \tau_a \lambda_b C\bar{q}^T)
	(q^T C i\gamma_5\tau_a\lambda_b\,q),
\label{qqlag}
\end{eqnarray}
where $\tau_a$ and $\lambda_b$ are the antisymmetric Gell-Mann matrices acting in, respectively, flavor and color space. $G_S$, $K$ and $G_D$ are coupling constants that must be determined by experiments.

The quark-quark interaction term, ${\cal L}_{qq}$, gives rise to superconducting condensates, $s_{ab}=\left<q^TC\gamma_5\tau_a\lambda_bq\right>$, which break SU(3)$_c$ and U(1) symmetry.
% A number of the gluons therefore acquire a mass through the Anderson-Higgs mechanism (Meissner effect). The condensates have net electric charge so the photon also acquires a mass and mixes with the gluons. 
The symmetries of $\cal L$ corresponds to a conserved chromo-electromagnetic charge.
The associated chemical potential is~\cite{Buballa:2003qv}
\begin{equation}
\hat{\mu} = \mu + \mu_Q\left(\frac{\tau_3}{2}+\frac{\tau_8}{2\sqrt{3}}\right) + \mu_3\lambda_3 + \mu_8\lambda_8.
\label{chempot}
\end{equation}
Here, $\mu$ is the quark-number chemical potential, $\mu_Q$ the positive electric-charge chemical potential, while $\mu_3$ and $\mu_8$ are color-charge chemical potentials. By linearizing \eqn{lagrangian} in the quark-quark (diquark) gaps, $\Delta_{ab} = 2G_Ds_{ab}$, and the quark-antiquark (chiral) gaps, $\phi_i = -4G_S\left<\bar{q}_iq_i\right>$, a grand canonical thermodynamic potential can be obtained by standard methods~\cite{Buballa:2003qv, Ruster:2005jc,Blaschke:2005uj,Abuki:2005ms}:
\begin{eqnarray}
	\Omega(T,\mu) &=& \frac{\phi^2_u+\phi^2_d+\phi^2_s}{8 G_S}
	+\frac{K\phi_u\phi_d\phi_s}{16G^3_S} \nonumber \\
	&+&\frac{\Delta^2_{ud}+\Delta^2_{us}+\Delta^2_{ds}}{4 G_D} \nonumber \\
	&-&\int\frac{d^3p}{(2\pi)^3}\sum_{n=1}^{18}
	\left[E_n+2T\ln\left(1+e^{-E_n/T}\right)\right] \nonumber \\
	&+& \Omega_{lep} - \Omega_0.
\label{potential}
\end{eqnarray}
Here, $E_n(p,\,\mu;\,\mu_Q,\mu_3,\mu_8,\,\phi_u,\phi_d,\phi_s,\,\Delta_{ud},\Delta_{us},\Delta_{ds})$ are the quasiparticle dispersion relations, $\Omega_{lep}$ is the contribution from leptons ({\it e.g.}, electrons, muons and the corresponding neutrino flavors) and $\Omega_0$ is the vacuum\ie $\Omega(0,\,0)=0$. It should be noted that~\eqn{potential} is an even function of $E_n$, so the signs of the dispersion relations are arbitrary. We therefore follow the standard convention that all states below the Fermi surface ($E_n<0$) are occupied, and only positive energy states are considered. In~\eqn{potential} the diquark gaps are denoted with flavor indices. This can readily be done by considering the color and flavor structure of the Gell-Mann matrices
\begin{eqnarray}
	\Delta_{ud} &\equiv \Delta_{22} &\;\;\;(u-d,\;r-g\;\text{pairing}), \\
	\Delta_{us} &\equiv \Delta_{55} &\;\;\;(u-s,\;r-b\;\text{pairing}), \\
	\Delta_{ds} &\equiv \Delta_{77} &\;\;\;(d-s,\;g-b\;\text{pairing}),
\end{eqnarray}
and $\Delta_{ab}=0$ if $a\neq b$~\cite{Buballa:2003qv}. The chiral gaps and the diquark gaps are variational parameters that are determined by minimization of~\eqn{potential}. The constituent quark masses are
\begin{equation}
	M_i = m_i + \phi_i + \frac{K}{8G^2_S}\phi_j\phi_k,
\label{constmass}
\end{equation}
where $(i,\,j,\,k)$ is any permutation of $(u,\,d,\,s)$.

In QCD, a color superconducting ground state is automatically color neutral due to the generation of gluon condensates in one or more of the eight components of the gluon field. In NJL models there are no gauge fields that neutralize the color charge dynamically, because the gluons have been replaced by effective point-like quark-antiquark~\eqn{qbarqlag} and quark-quark~\eqn{qqlag} interactions. Color neutrality must therefore be enforced by solving for the charge chemical potentials, $\mu_Q$, $\mu_3$ and $\mu_8$, such that the corresponding charge densities, $n_a=\left<\psi^\dagger T_a\psi\right>=-\partial\Omega/\partial\mu_a$, are zero~\cite{Buballa:2005bv}.

The values of the gaps and the (charge) chemical potentials depend on the coupling constants ($G_S$, $K$ and $G_D$), the current quark masses ($m_u$, $m_d$ and $m_s$) and the regularization method. These input parameters are fitted to low-density hadronic results and are therefore only approximately known. In addition, approximations are frequently used in order to simplify the evaluation of~\eqn{potential}. In this context it would be useful to have a mathematically exact condition for the apperance of gapless quasiparticle dispersion relations, without reference to specific input parameters and further assumptions. This condition is presented below.

The dispersion relations, $E_n$, are eigenvalues of six 4x4 matrices and one 12x12 matrix~\cite{Ruster:2005jc,Blaschke:2005uj,Abuki:2005ms}. Disregarding the signs, three 4x4 matrices and six of the twelve eigenvalues of the 12x12 matrix remain ($3\times4+6=18)$. The 12x12 matrix corresponds to $ur-dg-sb$ pairing and the three 4x4 matrices correspond to $ug-dr$, $ub-sr$ and $db-sg$ pairing. There are strong indications that the $ur-dg-sb$ modes are never gapless, because the Fermi momenta of these three species are approximately equal~\cite{Alford:2004hz}, and no such gapless modes have been found in numerical evaluations~\cite{Ruster:2005jc,Blaschke:2005uj,Abuki:2005ms}. A proof has turned out to be difficult to obtain due to the complexity of the characteristic polynomial of the 12x12 matrix. We therefore leave this analysis to a future publication. Here the 4x4 matrices are considered. The characteristic polynomials of these matrices can be written as
\begin{equation}
        E_n^4 + a_3 E_n^3 + a_2 E_n^2 + a_1 E_n + a_0.
\label{charpol}
\end{equation}
The $a_0$ coefficient of the polynomial is~\cite{Blaschke:2005uj}
\begin{eqnarray}
	&a_0 = p^4+\left(M^2_i+M^2_j+2\Delta^2_{ij}-\mu^2_{i\alpha}-\mu^2_{j\beta}\right)p^2& \nonumber \\
	&+\left(\mu_{i\alpha}\mu_{j\beta}+M_iM_j+\Delta^2_{ij}+\mu_{i\alpha}M_j+\mu_{j\beta}M_i\right)& \nonumber \\
	&\;\,\times\left(\mu_{i\alpha}\mu_{j\beta}+M_iM_j+\Delta^2_{ij}-\mu_{i\alpha}M_j-\mu_{j\beta}M_i\right),&
\label{a0coef}
\end{eqnarray}
for quark flavors $(i,\,j)$ and colors $(\alpha,\,\beta)$. The chemical potential, $\mu_{i\alpha}$, for a quark field with flavor $i$ and color $\alpha$ can be extracted from~\eqn{chempot} ($\hat{\mu}$ is diagonal in color and flavor space). A gapless dispersion relation is characterized by $E_n(p)=0$ for some real value(s) of $p$ when $\Delta_{ij}\neq0$. This requires that $a_0(p)=0$ has at least one real root. The solutions are
\begin{eqnarray}
	&&p^2=\bar{\mu}^2+{\delta\mu}^2-\bar{M}^2-{\delta M}^2-\Delta^2 \nonumber \\
	&&\pm2\sqrt{(\bar{\mu}\delta\mu-\bar{M}\delta M)^2-\Delta^2(\bar{\mu}^2-{\delta M}^2)}.
\label{roots}
\end{eqnarray}
Here we have introduced the quantities
\begin{eqnarray}
	&\bar{M} = (M_i+M_j)/2,
	&\;\;\;\;\delta M = (M_i-M_j)/2, \label{avdmass} \\
	&\bar{\mu} = (\mu_{i\alpha}+\mu_{j\beta})/2,
	&\;\;\;\;\delta \mu = (\mu_{i\alpha}-\mu_{j\beta})/2, \label{avdmu}
\end{eqnarray}
and $\Delta=\Delta_{ij}$. The indices in~\eqn{a0coef} can be omitted without ambiguity, since we are dealing with two-species pairing. Observe that the masses and chemical potentials of the paired quark fields are $\bar{M}\pm\delta M$ and $\bar{\mu}\pm\delta\mu$. A real square root in \eqn{roots} requires that
\begin{equation}
	\Delta \leq \Delta^g \equiv \displaystyle \frac{|\bar{\mu}\delta\mu - \bar{M}\delta M|}
	{\sqrt{\bar{\mu}^2-\delta M^2}},
\label{gapless4x4_1}
\end{equation}
and a positive solution for $p^2$ requires that
\begin{eqnarray}
	&&\bar{M}^2 + {\delta M}^2 + \Delta^2 - \bar{\mu}^2 - {\delta\mu}^2 \nonumber \\
	&&\leq2\sqrt{(\bar{\mu}\delta\mu-\bar{M}\delta M)^2-\Delta^2(\bar{\mu}^2-{\delta M}^2)}.
\label{gapless4x4_2}
\end{eqnarray}

\begin{figure*}
\epsfig{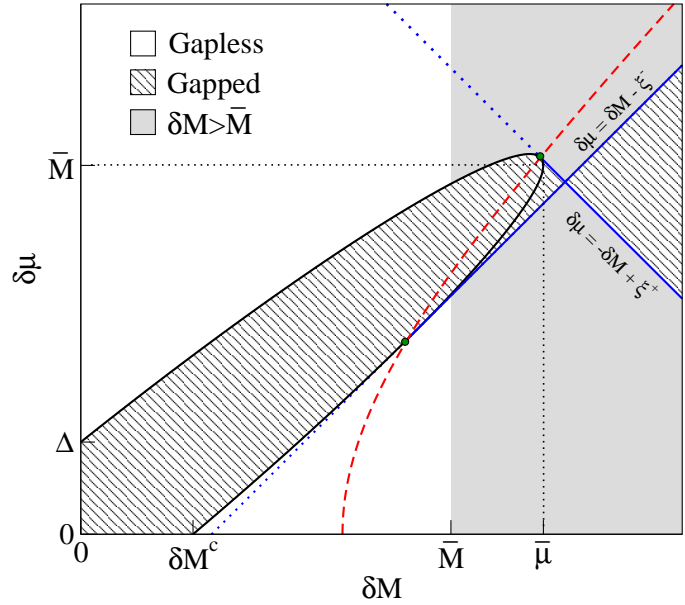}
\caption{Graphical representation of the gapless inequalities~\eqns{gapless4x4_1}{gapless4x4_2}. For clarity only the first quadrant is shown. The third quadrant is a reflection of the first quadrant in the origin. In this figure an unreasonably large value of $\bar{M}$ has been used in order to emphasize the role of the tangent lines. Qualitatively, the gapped region can be represented by the rotated ellipse, see the text. The values of $\delta\mu$ and $\delta M$ can be represented by a point in the $\delta M-\delta \mu$ plane. If this point is enclosed by the hatched area the dispersion relations are gapped. Otherwise a gapless dispersion relation exists. Here $\delta M^c=\bar{\mu}/(1+\bar{M}^2/\Delta^2)^{1/2}$ and $\xi^\pm=[\Delta^2+(\bar{\mu}\pm\bar{M})^2]^{1/2}$.}
\label{ellipsefig}
\end{figure*}

Inequality~\eqn{gapless4x4_1} can be represented with a rotated ellipse in the $\delta M-\delta\mu$ plane, as in \fig{ellipsefig}. The interior region of the ellipse violates~\eqn{gapless4x4_1} and hence represent gapped modes. Outside the ellipse the square root in~\eqn{gapless4x4_2} is real, and~\eqn{gapless4x4_2} is obviously satisfied as long as the left-hand side is negative. A negative left-hand side of~\eqn{gapless4x4_2} is represented by the region in-between the two branches of the hyperbola, $\bar{\mu}^2+{\delta\mu}^2=\bar{M}^2+{\delta M}^2+\Delta^2$, in \fig{ellipsefig}. For a positive left-hand side, which corresponds to the two regions on the left- and right-hand side of the hyperbola,~\eqn{gapless4x4_2} can be squared, and four coupled inequalities linear in $\delta M$ and $\delta\mu$ are obtained. These correspond to tangent lines of the ellipse. The hatched areas enclosed by the tangent lines, the hyperbola and the ellipse violate~\eqn{gapless4x4_2} and hence represent gapped modes. For each tangent line the intersection with the hyperbola coincides with the point on the ellipse. Inequality~\eqn{gapless4x4_2} is relevant if $\delta M/\bar{\mu}\sim 1-\Delta^2(\bar{\mu}-\bar{M})^{-2}/2$, which is not the case for realistic values of the masses and chemical potentials. This is explicitely demonstrated by the examples in \fig{dispelfig}. Inequality~\eqn{gapless4x4_1} is therefore the relevant condition for gapless modes. In achieving this result, no further approximations than those leading up to~\eqn{potential} were made.

For the two-flavor color superconducting (2SC) phase, which is characterised by $\Delta_{ud}\neq0$ and $\Delta_{us}=\Delta_{ds}=\mu_3=0$, one can use the fact that $\delta M \ll \delta\mu$ and $M\ll\mu$, so the gapless condition~\eqn{gapless4x4_1} is approximately
\begin{equation}
	\Delta_{ud} \lesssim |\delta\mu|=-\mu_Q.
\label{gapless2SC}
\end{equation}
For the three-flavor color-flavor locked (CFL) phase, which is characterised by $\Delta_{ij}\neq 0$ and $\bar{M}\sim\delta M\sim M_s/2$, a series expansion of~\eqn{gapless4x4_1} to first order in $M_s^2/\bar{\mu}$ yields
\begin{equation}
	\Delta_{is} \lesssim |\delta\mu|+\left[\frac{\text{sign}(\delta\mu)}{4}+\frac{|\delta\mu|}{8\bar{\mu}}\right]\frac{M^2_s}{\bar{\mu}},
\label{gaplessCFL}
\end{equation}
where $\delta\mu=(\mu_{i\alpha}-\mu_{s\beta})/2$. These well-known approximative results are instructive at the qualitative level, but should not be used mechanically. See discussion below.

\begin{figure*}
\epsfig{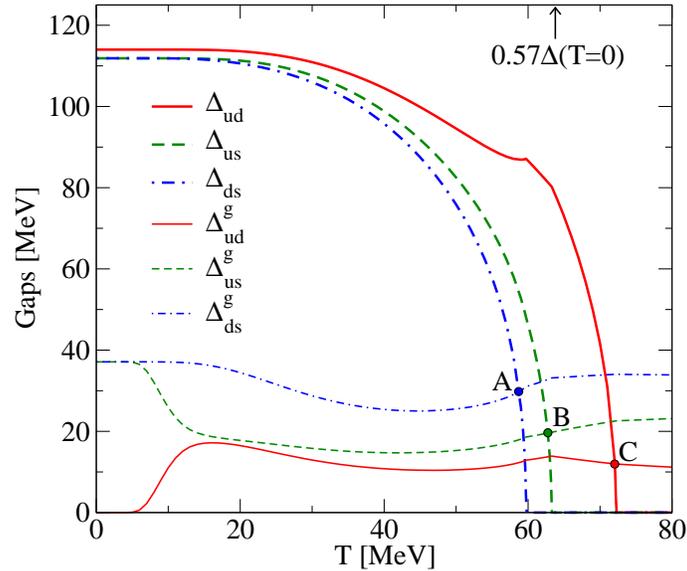}
\caption{Diquark gaps vs. the temperature at $\mu=500$~MeV and $\eta=0.75$. $\Delta^g_{ij}$ is the threshold
for gapless quasiparticle dispersion relations~\eqn{gapless4x4_1}\ie gapless modes exist iff $\Delta_{ij}\leq\Delta^g_{ij}$. The critical points where gapless $db-sg$, $ub-sr$ and $ug-dr$ quasiparticles appear are denoted with, respectively, $A$, $B$ and $C$. The BCS result for the critical temperature of a superconducting condensate, $T\sim 0.57\Delta(T=0)$, is indicated in the plot. This figure represents a cross-section of \figN{5} in~\cite{Blaschke:2005uj}.}
\label{gapfig}
\end{figure*}

\begin{figure*}
\epsfig{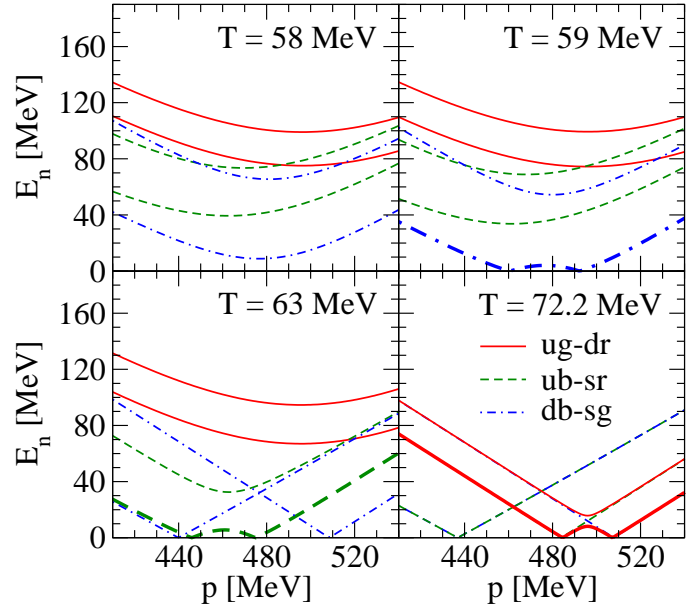}
\caption{Dispersion relations at $\mu=500$~MeV and $\eta=0.75$ for four different temperatures. Gapless modes are denoted with bold lines. Gapped modes and modes of unpaired quarks are denoted with thin lines. Compare with \fig{gapfig}.}
\label{dispfig}
\end{figure*}

\begin{figure*}
\epsfig{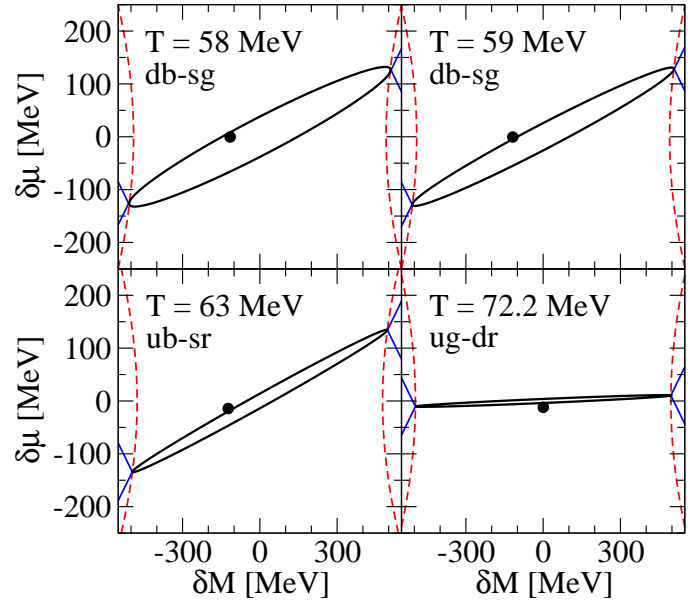}
\caption{Graphical representation of the gapless conditions~\eqns{gapless4x4_1}{gapless4x4_2} for some of the quasiparticle dispersion relations represented in \fig{dispfig}. The values of $\delta M$ and $\delta\mu$ are represented by bold points. If the center of a point is enclosed by the ellipse the corresponding quasiparticle have gapped dispersion relations, otherwise a gapless dispersion relation exists.}
\label{dispelfig}
\end{figure*}

Next we present a numerical example and therefore constrain the discussion to a specific parametrization of the model, as in~\cite{Blaschke:2005uj}. The momentum integral is regularized with a cut-off, $\Lambda=602.3$~MeV. The coupling constants are $G_S\Lambda^2=2.319$, $G_D/G_S\equiv\eta=0.75$ and $K=0$. The current quark masses are $m_u=m_d=5.5$~MeV and $m_s=112$~MeV. Inserting these parameters in the thermodynamic potential~\eqn{potential}, the gaps ($\phi_u,\phi_d,\phi_s,\,\Delta_{ud},\Delta_{us},\Delta_{ds}$) can be determined by minimization of the free energy, while simultaneously neutralizing all charge densities with $\mu_Q$, $\mu_3$ and $\mu_8$. In \fig{gapfig} the diquark gaps, $\Delta_{ij}$, and the gapless thresholds~\eqn{gapless4x4_1}, $\Delta^g_{ij}$, are plotted vs. the temperature for a fixed value of the quark number chemical potential, $\mu=500$~MeV. In \fig{dispfig} the quark-quark quasiparticle dispersion relations are plotted for four different temperatures represented in \fig{gapfig}. Observe that gapless dispersion relations exist iff $\Delta_{ij}\leq\Delta^g_{ij}$. \fig{dispelfig} shows the graphical representation of the gapless condition for some quasiparticles represented in \fig{dispfig}, see \fig{ellipsefig} for further information.

We find that the difference between the approximative results ~\eqns{gapless2SC}{gaplessCFL} and the exact gapless condition~\eqn{gapless4x4_1} is typically below 5\% in the plane of temperature and quark number chemical potential. However, even a small error in $\Delta^g_{ij}$ could lead to qualitatively incorrect conclusions if $\Delta_{ij}(T\sim 0)\sim\Delta^g_{ij}$, because $\Delta_{ij}(T)$ are roughly constant at the low temperatures relevant for compact star evolution. The exact condition for gapless quasiparticle excitations presented here~\eqn{gapless4x4_1}, which is the main result of this paper, is a safe alternative to the approximative results. Moreover, \fig{ellipsefig} is an accurate qualitative picture of the prerequisites for gapless color-$\bar{3}$ excitations in NJL models.

\begin{acknowledgments}
We thank D. Blaschke, M. Buballa, S. Fredriksson and H. Grigorian for useful discussions and suggestions. A.M.\"O. received support from Hacettepe University Research Fund, grant No. 02 02 602 001. F.S. acknowledges support from the Swedish National Graduate School of Space Technology and thanks D. Blaschke and the Organizers of the Helmholtz International Summer School, Dubna, for partial support and their hospitality.
\end{acknowledgments}

%%%%%%%%%%%%%%%%%%%%%%%%%%%%%%%%%%%%%%%%%%%%%%%%%%%%%%%%%%%%%%%%%%%

\bibliographystyle{h-physrev.bst}
\bibliography{gapless.bib}

%%%%%%%%%%%%%%%%%%%%%%%%%%%%%%%%%%%%%%%%%%%%%%%%%%%%%%%%%%%%%%%%%%%

\clearpage

\end{document}